\documentclass{pasj00}

\begin{document}
\SetRunningHead{S. Yuki \& S. Shibata}{A Particle Simulation for the
Pulsar Magnetosphere}
\Received{2011 July 14}\Accepted{2011 November 8}\Published{2012 June 25}
\title{A Particle Simulation for the Pulsar Magnetosphere: Relationship
of Polar Cap, Slot Gap, and Outer Gap}

\author{Shinya \textsc{Yuki} and Shinpei \textsc{Shibata} }
\affil{Graduate School of Science and Engineering, Yamagata University

1-4-12 Kojirakawa, Yamagata 990-8560}
\email{yuki@ksirius.kj.yamagata-u.ac.jp, shibata@ksirius.kj.yamagata-u.ac.jp}

\KeyWords{magnetic fields -- plasmas -- pulsars: general.} 
\maketitle

\begin{abstract}
To explain the pulsed emission of the rotation powered pulsars from radio
to gamma-ray, the polar cap models, the slot gap models,
and the outer gap models are proposed. The recent observations suggest that
these models are likely to co-exist in the same magnetosphere. If so, their
mutual relation is known to be troublesome \citep{b25} due to
the  boundary conditions and the direction of the current which are properly
assumed in each acceleration models.

We performed a particle simulation for the global magnetospheric structure.
Based on the simulation, we present a new picture of the global structure
of the pulsar magnetosphere.
It is found that a new dead zone is formed along the current neutral line which
separates the oppositely directed current. We shall call this the current-
neutral zone.  We suggest that the polar cap accelerators and the slot gaps
locate above the current-neutral zone, and the outer gap exist between the
current neutral zone and the traditional dead zone.
We also give an estimate of the super-rotation region.
\end{abstract}

\section{Introduction}
The mechanism of the pulsed emission from pulsars has been an important
unsolved problem for more than forty years.
The recent observations of the $\textit{Fermi}$ Gamma-ray Space Telescope
has promoted an understanding of the emission from gamma-ray pulsars.
The pulse shapes show substantial diversity, but roughly 75$\%$ of the gamma-ray profiles have two peaks,
separated by $\gtrsim 0.2$ of rotation phase \citep{b1}. 
Additionally,  most of the radio-loud gamma-ray pulsars have the first
gamma-ray peak lagging the main radio component.
This indicates that radio and gamma-ray beams are coming from different emitting regions.
Also, the high-confidence gamma-ray from a millisecond pulsar is detected first by $\textit{Fermi}$. After that,
it was found that most of the millisecond pulsars have the features of the gamma-ray light curves and spectra like
normal pulsars.

Polar cap (PC) (eg, \cite{b2}),
 slot gap (SG) (eg, \cite{b3,b4}) and
 outer gap (OG) (eg, \cite{b5,b6}) models
 are promising particle acceleration models to explain pulsed emission from pulsars.
In PC models, particle acceleration and resultant gamma-ray emission takes place in the open field line region above the magnetic poles.
The gamma-rays undergo the magnetic pair production, which leads to super-exponential spectral cutoff.
Then, the produced pairs screen out the accelerating electric field above the so-called pair formation front (PFF).
It is generally believed that the observed radio emission is basically
 coming from upper part of the PC region. 
The SG, which is the rim of the PC, extends along the last-open-field line.
There is no sufficient acceleration to emit gamma-rays producing electron-positron pairs at the low altitude
because the field-aligned electric field near the last open field line is weak.
Therefore, particle acceleration lasts up to high altitudes of many stellar radii.
For this reason, the SG can provide extended high-energy emission above the last-open-field lines.
The OG is located between a neighborhood of the null charge surface and the light cylinder along the open field lines.
The accelerating electric field exists because of charge depletion, and photon-photon pair production occurs in the gap.
The OG can also provide extended high energy emission.
In addition, SG and OG models have roughly simple exponential spectral
 cutoff, which is more gradual than that of PC models.

SG and/or OG models can explain the double-peaked gamma-ray light curves
(eg, \cite{b7,b8}).
On the other hand, PC models have much more difficulty reproducing the observed wide pulse profile because the beam size from the PC is too small.
Furthermore, the possibility of gamma-ray emission from the PC, which would exhibit super-exponential cutoff due to magnetic pair attenuation, was
eliminated by the spectral observation such as MAGIC and
$\textit{Fermi}$ (eg, \cite{b9,b1}).
Also, there are not only the study of light curves and averaged spectrum
but also that of phase-resolved spectra (eg, \cite{b10,b11}).

In PC, SG and OG models, acceleration is caused by an electric field parallel to the magnetic field line,
and the accelerating electric field is calculated by the Poisson equation for the non-corotational potential:
\begin{equation}
- \nabla^2 \Phi = 4 \pi (\rho - \rho_{\mathrm{GJ}}),
\end{equation}
where $\boldsymbol{E} = -(\boldsymbol{\Omega} \times \boldsymbol{r})
\times \boldsymbol{B}/c - \nabla \Phi$; $\rho$ and
$\rho_{\mathrm{GJ}} = - \boldsymbol{\Omega} \cdot \boldsymbol{B} /
(2 \pi c) [1 - (\boldsymbol{\Omega} \times \boldsymbol{r} / c)^2]$
are the charge density and Goldreich-Julian charge density \citep{b12}, respectively; $\boldsymbol{\Omega}$ is the angular momentum; and
$\boldsymbol{r}$ is the position vector.
Thus, the field-aligned electric field $E_{||} = -\nabla \Phi \cdot \hat{\boldsymbol{B}}$,
with $\hat{\boldsymbol{B}}$ unit vector along the magnetic field line,
is formed by the difference between
local charge density and Goldreich-Julian density.

The boundary conditions of these models, which are required to solve the Poisson equation, are summarized as follows.
Most of the PC and SG models assume $E_{||}=0$, as well as $\Phi=0$ at
the stellar surface, and $\Phi=0$ along the last open field line.
In addition, the condition $E_{||}=0$ is imposed at an upper boundary,
ie, PFF.
The OG models, which elongate along open field lines, usually have upper and lower boundaries along the field lines; an inner boundary, which is on the star side; and an outer boundary, which is located near the light cylinder.
In particular, the lower boundary is supposed to be the last open field line.
A standard boundary condition is such that $\Phi=0$ on the upper and lower boundaries, $\Phi=0$ and $E_{||}=0$ on the inner boundary,
and $E_{||}=0$ on the outer boundary.
The boundary conditions are very important because they affect the structure of
the accelerating electric field.
However, the screening process on each boundary has not been well modeled or simulated.
Therefore, it is not clear whether $\Phi=0$ along the last open field line or not.

There is also a problem with the directions of the current flow.
The directions of the current along the magnetic field line above the last open field line
are different in these acceleration models.
When $\boldsymbol{\mu} \cdot \boldsymbol{\Omega} > 0$, where $\boldsymbol{\mu}$ is the magnetic moment,
the direction of the current in the PC and SG is inward, and while the current in the OG is outward.
When $\boldsymbol{\mu} \cdot \boldsymbol{\Omega} < 0$, the directions
are reversed.
Looking at various pulse shapes from radio to gamma-rays, people
sometimes speculate all these types of acceleration co-exist.
If all of these acceleration models exist above the same last open field line, the directions of the currents in these models are opposite.
Therefore, it seems to be impossible that polar-slot gap and OG exist above the same last open field lines consistently from the theoretical prospective.

On the other hand, most of the observed radio and gamma-ray features
of both normal and millisecond pulsars are well reproduced by radio
emission from upper part of the PC region and gamma-ray
emission from the SG or OG (eg, \cite{b13,b8}).
It is required on the observational side that the PC and SG and/or OG co-exist.

To understand the formation mechanism of the PC, SG and OG in view of electrodynamics,
particle simulation for the global magnetosphere would be a powerful way.
By using GRAPE \citep{b14}, a massively-parallel
special-purpose computer for astronomical N-body simulations at National
Astronomical Observatory of Japan,
we performed a simulation particularly aimed at understanding the boundary conditions for the PC, SG and OG; the relationship among the models; and the current directions.

The organization of this paper is as follows:
Section 2 describes the method of our simulation; Section 3 shows the simulation results and describes the structure of the global pulsar magnetosphere; Section 4 discusses the position relationships of the PC, SG, and OG, and summarizes the results of our simulation. Our main conclusion, that both polar-slot gap and OG can exist in a magnetosphere, is also presented in Section 4.

\section{Method of Simulation}
It is necessary to handle formation of the field-aligned electric field $E_{||}$,
so that the simulation must not be the usual MHD simulation.
A breakthrough can be found by a global particle simulation, which can handle cross-field drift motions due to
radiation drag and particle inertia by solving each particle motion.
\authorcite{b15} (\yearcite{b15}, hereafter WS) have succeeded in overcoming this heavy computation by using GRAPE.
Typical use of GRAPE is for stellar dynamics.
However, it should be noted that GRAPE, in which sign-bit is available, can calculate Coulomb force as well as gravitational force because both of them are represented by the inverse-square law:
\begin{equation}
\boldsymbol{E}_i = - \sum_{j \neq i}^N q_j \frac{\boldsymbol{r}_j - \boldsymbol{r}_i}{( |\boldsymbol{r}_j - \boldsymbol{r}_i|^2 + \epsilon^2)^{3/2}},
\end{equation}
where the subscripts $i$, $j$ represent $i/j$th particle;
and $N$,$\boldsymbol{r}$, $\epsilon$, $q$, and $\boldsymbol{E}$
are the total particle number, position vector, softening parameter,
charge, and electric field, respectively.

In our simulation, we assumed that a star is an aligned rotator and a
perfect conductor.
To specify the charge sign, we chose aligned ($\boldsymbol{\mu} \cdot
\boldsymbol{\Omega} > 0$), not counter aligned ($\boldsymbol{\mu} \cdot
\boldsymbol{\Omega} < 0$), ie, the poles are negatively charged, while
the equator is positively charged.
The numerical method follows WS.
Magnetospheric plasmas are represented as super-particles.
The electric field on the $i$th particle is calculated by the superposition of the
vacuum solution $\boldsymbol{E}_{\mathrm v}$ and the space-charge solution $\boldsymbol{E}_{\mathrm q}$, where
\begin{eqnarray}
\boldsymbol{E}_{\mathrm v} (\boldsymbol{r}_i) = - \left[ \frac{\Omega \mu R^2 (3\cos^2 \theta -1)}{c r^4_i} - \frac{Q_{\mathrm {sys}}}{r^2_i} \right] \boldsymbol{e}_{\mathrm r} \nonumber
\\ - \left( \frac{2 \Omega \mu R^2 \sin \theta \cos \theta }{c r^4_i}  \right) \boldsymbol{e}_{\mathrm \theta},
\end{eqnarray}
\begin{eqnarray}
\boldsymbol{E}_{\mathrm q}  (\boldsymbol{r}_i) = \sum_{j \neq i}^N  \biggl[
\frac{\boldsymbol{r}_i - \boldsymbol{r}_j }{|\boldsymbol{r}_i - \boldsymbol{r}_j|^3}
- \frac{R}{r_j} \frac{\boldsymbol{r}_i - (R/r_j)^2 \boldsymbol{r}_j
}{|\boldsymbol{r}_i - (R/r_j)^2 \boldsymbol{r}_j|^3} \nonumber
\\ - \left( 1 - \frac{R}{r_j} \right) \frac{\boldsymbol{r}_i}{r^3_i}  \biggr],  \label{eq:01}
\end{eqnarray}
where $\boldsymbol{e}_{\mathrm r}$ and $\boldsymbol{e}_{\mathrm \theta}$ are unit vectors in spherical coordinates;
$R$, $Q_{\mathrm {sys}}$, and $c$ are the stellar radius, system charge
and speed of light, respectively.
To satisfy the boundary condition for a perfect conductor, the second
and third terms  in (\ref{eq:01}) representing the effect of mirror charge exist.

As an improvement from WS, we take into account the modification of the magnetic field from the dipole field.
By using GRAPE, the magnetic field is calculated by the Biot-Savart law:
\begin{eqnarray}
\boldsymbol{B} (\boldsymbol{r}_i) = 
B_* \left( \frac{R}{r_i} \right)^3 (\cos \theta  \boldsymbol{e}_{\mathrm r} + \frac{1}{2} \sin \theta \boldsymbol{e}_{\mathrm \theta})
\\ - \sum_{j \neq i}^N \frac{q_j \boldsymbol{v}_j \times  (\boldsymbol{r}_j - \boldsymbol{r}_i)}{( |\boldsymbol{r}_j - \boldsymbol{r}_i|^2 + \epsilon^2)^{3/2}},
\end{eqnarray}
where $\boldsymbol{v}_j$ is the velocity of the $j$th particle.

We made the following assumptions as WS did:
(1) particles are emitted freely from the stellar surface;
(2) the particles are subject to the radiation drag force;
and (3) pair creation occurs where the field-aligned
electric field $E_{||}$ is stronger than a critical value $E_{\mathrm {cr}}$.
Since we are looking for an axisymmetric steady state, we ignored the effect of
the time variation of the fields.
Therefore, the problem is
electro-static and magneto-static, and whereby GRAPE can be used.
Our simulation is as follows: starting from vacuum around the star, the surface charge is pulled out from the
star and, as it spreads throughout the magnetosphere, radiation drag
takes place; and pairs are created, if $E_{||} > E_{\mathrm {cr}}$.
This simulation proceeds in the following steps:
\begin{enumerate}
 \item Start the calculation from the vacuum around the star.

 \item Replace the surface charges on the star by particles.

 \item Compute the electromagnetic field at the particles' positions.

 \item Compute the force exerted on the particles, and advance the particle positions and velocities by the equations of motion.

 \item Create electron-positron pairs where $E_{||} > E_{\mathrm {cr}}$.

 \item Remove particles which fall back to the star or leave the simulation box through the outer boundary.

 \item Go back to step (2) unless the steady state is established.
\end{enumerate}
Our simulation is started from the vacuum state.
Then, the surface charges are pulled out by the induced electromotive force which is much larger than the work function.
They are emitted as charged particles above the stellar surface.
The emission is expected to last unless $E_{||} \neq 0$.
Therefore, we emitted the surface charge $\sigma$ per unit area which is defined as
$E_{||} = 4 \pi (\sigma - \sigma_{\mathrm {GJ}}) (\boldsymbol{n} \cdot \hat{ \boldsymbol{B}})$,
where $\sigma_{\mathrm {GJ}}$ is the surface charge density when the
corotational electric field exists both inside and outside of the star;
$\boldsymbol{n}$ is the unit normal vector to the stellar surface; and $\hat{ \boldsymbol{B}}$ is the unit vector of the magnetic field on the stellar surface.
The surface charges are emitted $\sim 230$ times in a rotation period.

We solve the equation of motion, 
\begin{equation}
 \frac{{\mathrm {d}} \boldsymbol{p}_i}{{\mathrm {d}} t}
 = 
 q_i \left( \boldsymbol{E} + \boldsymbol{\beta}_i
 \times \boldsymbol{B} \right)
 + \boldsymbol{F}_{\mathrm {rad}},
\end{equation}
where $\boldsymbol{\beta}_i$ is the velocity of the $i$th particle in units of
the speed of light, $\boldsymbol{p}_i = \gamma_i m_i \beta_i c$, and $\gamma_i = (1 - \beta_i^2)^{-1/2}$;
$\boldsymbol{F}_{\mathrm {rad}} = (2/3) (q_i^2 / R_{\mathrm c}^2) \gamma_i^4
(\boldsymbol{p}_i / |\boldsymbol{p}_i|)$ (if $\beta_i \simeq 1$) is the
radiation reaction force, and $R_{\mathrm c}$ is the curvature radius.
This approximate expression for the radiation drag force can be used when the force due to the external field is larger than the radiation drag force.

For pair creation, grid points $\boldsymbol{r}_{i,j} = (r_i, \theta_j)$ are provided on the meridional plane.
If $E_{||}$ is larger than the critical value $E_{\mathrm {cr}}$, $\alpha |E_{||}(\boldsymbol{r}_{i,j})| / E_{\mathrm {cr}}$ pairs whose initial velocities are zero
are generated in the grid spaces.
Here the multiplicity $\alpha$ is set to $\alpha = 2$ and $E_{\mathrm {cr}} = 0.25 B_{\mathrm L}$, where $B_{\mathrm L} = \mu/R_{\mathrm L}^3$
is the light cylinder magnetic field, in the present simulation.

The volume of the grid spaces is in proportion to $r^{-2}$, so that pair creation rate per unit volume is in proportion to $r^{^2}$.
This reproduces the fact that the density of target photons supplied from the stellar surface decreases in proportion to $r^{2}$.
A time interval for the pair creation is also the parameter of the calculation.
In the present calculation, it is such that pairs are created $\sim 75$ times in a rotation period.

The outer boundary is located at 10 $R_{\mathrm L}$.
Magnetospheric plasmas are expressed as super-particles, and therefore the exaggerated electric force causes a bound motion like positronium.
We remove such bound pairs because they behave as if they were a neutral particle, and they would not affect the structure of the magnetosphere. 
In the steady state, creation and loss of the particles in the simulation balance with each other.
Thus, the total charge of the system is determined automatically.

As for normalization, the position vector, magnetic field, electric field, and velocity are normalized as follows:
$\bar{ \boldsymbol{r} } = \boldsymbol{r} / R$, $\bar{ \boldsymbol{B} } = \boldsymbol{B} / B_*$, $\bar{ \boldsymbol{E} } = \boldsymbol{E} / B_*$, and
$\boldsymbol{\beta} = \boldsymbol{v} / c$,
where $B_*$ is the magnetic field strength on the poles;
the variables with bar are non-dimensional in this paper. 
In our simulation, the strength of the magnetic field is normalized by
$B_*$, so that our simulation can treat pulsars which have various
magnetic field strengths basically.
The other major simulation parameters are
$\bar{ \Omega } = R/R_{\mathrm L} =
0.2$, 
$\bar{m} = 10^{-10}$,
and $\bar{q} = 10^{-5}$,
where $R_{\mathrm L}$ is the light cylinder radius;
$\bar{m}$ and $\bar{q}$ are the mass and charge of a super-particle,
respectively, in non-dimensional form.

The simulation parameters should be chosen
so that what takes place in the simulation reproduces the actual phenomena
in the pulsar magnetosphere. 
At the same time, we have a restriction of the computational power.
Our parameters include the mass and charge of
the super-particles $\bar{m}$ and $\bar{q}$,
time steps of integration, 
angular velocity $\bar{\Omega}$, frequency of pair creation,
critical electric field for pair creation $E_{\mathrm {cr}}$, 
pair multiplicity $\alpha$, frequency of injection from the stellar surface, and
softening parameter of Coulomb force. It is notable that the total number
of particles in the simulation box is determined self-consistently but can
be controlled by $\bar{q}$ and
the parameters of pair creation.
We do not have any grid
points for the electromagnetic field.
We give typical scales of time and length
for the simulation, normal pulsars, and MSPs
in Table \ref{tab:00} and
\ref{tab:01},  where $\gamma_1$ and $\gamma_2$ are the Lorentz
factors in units of $10$ and $10^2$, respectively; $P_{-3}$ and $P_{1}$
are the stellar rotation
periods in units of $10^{-3}$ sec and $1$ sec,
respectively; $B_{8}$ and $B_{12}$ are the magnetic field strength in
units of $10^{8}$ G and $10^{12}$ G, respectively;
$R_6$ is the stellar radius in units of $10^6$ cm; and $M_2$
is the multiplicity which is defined as $M \equiv n / n_{\mathrm {GJ}}$ in
units of $10^2$.
 The time step was taken to be much smaller than the Larmor period to
 reproduce drift motions due to the electric field, the gradient of the
 magnetic field, the centrifugal force, and the radiation drag, and so
 on as seen in Table \ref{tab:00}. 
To consider the screening effect and the formation of the field-aligned
electric field due to inertia, the time step also has to be much smaller
than the the period of plasma oscillation. 
This requirement is also satisfied as seen in  Table \ref{tab:00}.
According to Table 2, the Larmor radii and inertial lengths are enough smaller
than the typical scale of the system.
The ratio of the plasma frequency  $\omega_p$ to the gyro frequency
$\omega_g$ changes in such a way that
$\omega_p / \omega_g \ll 1$  at the magnetic poles,
$\omega_p / \omega_g < 1$ at the light cylinder, and  
$\omega_p / \omega_g > 1$ at the outer boundary
in both simulation and actual pulsars.
Thus, the magnetization property of plasma is 
also simulated well in the simulation and 
actual pulsars.

\begin{table*}[]
  \caption{Typical periods for the simulation, normal pulsars, and
 MSPs.}
 \label{tab:00}
 \begin{center}
  \scalebox{0.9}{
   \begin{tabular}{c c c c}
    \noalign{\hrule height 1pt}
    & Simulation & Normal Pulsars (sec) & MSPs (sec)\\
    Time Step  & $10^{-5}$ & - & - \\  
    Period of Emitting Particles & $1.4 \times 10^{-1}$ & - & - \\
    Period of Generating Pairs & $4.2 \times 10^{-1}$ &  - & -\\
    Period of Rotation & $3.1 \times 10$  &  - & -\\
    Plasma Period at Pole  & $3.1 \times 10^{-2}$ & 
	    $4.2 \times 10^{-11} P_1^{1/2} 
	    B_{12}^{-1/2} M_2^{-1/2}$ & 
	    $1.3 \times 10^{-10} P_{-3}^{1/2} 
	    B_8^{-1/2} M_2^{-1/2}$\\
    Larmor Period at Pole  & $6.3 \times 10^{-4} \gamma_1$ &
	    $3.6 \times 10^{-17} \gamma_2 
	    B_{12}^{-1}$&
	    $3.6 \times 10^{-16} \gamma_2 
	    B_8^{-1}$ \\
    Plasma Period at LC & $5.0 \times 10^{-1}$ & 
	    $2.0 \times 10^{-5} P_1^{2} 
	    B_{12}^{-1/2} 
	    R_6^{-3/2}
	    M_2^{-1/2}$ & 
	    $2.0 \times 10^{-9} P_{-3}^{2} 
	   B_{8}^{-1/2} 
	   R_6^{-3/2}
	   M_2^{-1/2}$ \\
    Larmor Period at LC  & $1.6 \times 10^{-1} \gamma_1$ &
	    $7.8 \times 10^{-6} \gamma_2  P_1^{3}
	    B_{12}^{-1}
	    R_6^{-3}$&
	    $7.8 \times 10^{-11} \gamma_2  P_{-3}^{3}
	    B_8^{-1}
	   R_6^{-3}$\\
    Plasma Period at OB  & $5.0 \times 10$ & 
	      $2.0 \times 10^{-4} P_1^{2} 
	    B_{12}^{-1/2} 
	    R_6^{-3/2}
	    M_2^{-1/2}$ & 
	      $2.0 \times 10^{-8} P_{-3}^{2} 
	    B_8^{-1/2} 
	   R_6^{-3/2}
	    M_2^{-1/2}$ \\
    Larmor Period at OB  & $1.6 \times 10^2 \gamma_1$ &
	    $7.8 \times 10^{-3} \gamma_2  P_1^{3}
	    B_{12}^{-1} 
	    R_6^{-3}$&
	    $7.8 \times 10^{-8} \gamma_2  P_{-3}^{3}
	    B_8^{-1} 
	   R_6^{-3}$\\
    \noalign{\hrule height 1pt}
   \end{tabular}
  }
 \end{center}
 \end{table*}

\begin{table*}[]
  \caption{Typical lengths for the simulation, normal pulsars, and
 MSPs.}
 \label{tab:01}
  \begin{center}
   \scalebox{0.9}{
   \begin{tabular}{c c c c}
    \noalign{\hrule height 1pt}
      & Simulation & Normal Pulsars (cm) & MSPs (cm) \\
    Stellar Radius & 1 & $10^6$ & $10^6$\\
    Light Cylinder Radius & 5 & $4.8 \times 10^9 P_1$ & $4.8 \times 10^6 P_{-3}$\\
    Larmor Radius at Pole & $10^{-4} \gamma_1$ & 
	    $1.7 \times 10^{-7} \gamma_2 B_{12}^{-1}$& 
	    $1.7 \times 10^{-3} \gamma_2 B_8^{-1}$\\
    Inertial Length at Pole & $5 \times 10^{-3}$ &
	     $2.0 \times 10^{-1} P_1^{1/2} 
	    B_{12}^{-1/2} M_2^{-1/2}$&
	    $6.3 \times 10^{-1} P_{-3}^{1/2} 
	    B_8^{-1/2} M_2^{-1/2}$\\
    Larmor Radius at LC &  $2.5 \times 10^{-2} \gamma_1$ & 
	    $3.7 \times 10^4 \gamma_2 P_1^3
	    B_{12}^{-1}
	    R_6^{-3/2}$&
	    $3.7 \times 10^{-3} \gamma_2 P_{-3}^3
	    B_8^{-1} 
	R_6^{-3/2}$\\
    Inertial Length at LC & $7.9 \times 10^{-2}$ &
	     $9.4 \times 10^{4} P_1^{2} 	  
	    B_{12}^{-1/2}   R_6^{-3/2} M_2^{-1/2}$&
	     $9.4  P_{-3}^{2} 
	    B_8^{-1/2}  R_6^{-3/2} M_2^{-1/2}$ \\
    Larmor Radius at OB & $2.5 \times 10 \gamma_1$ &
	    $3.7 \times 10^7 \gamma_2 P_1^3	  
	    B_{12}^{-1}  R_6^{-3/2}$&
	    $3.7 \times 10^2 \gamma_2 P_{-3}^3	   
	    B_8^{-1}  R_6^{-3/2}$\\
    Inertial Length at OB & $7.9 \times 10^{-3}$ &
	    $9.4 \times 10^{3} P_1^{2} 	  
	    B_{12}^{-1/2}   R_6^{-3/2} M_2^{-1/2}$ &
	    $9.4 \times 10^{-1} P_{-3}^{2} 	  
	    B_8^{-1/2}   R_6^{-3/2} M_2^{-1/2}$\\
    \noalign{\hrule height 1pt}
   \end{tabular}
   }
  \end{center}
 \end{table*}

\section{Results}
\begin{figure*}
  \begin{center}
    \begin{tabular}{cc}
     \FigureFile(80mm,80mm){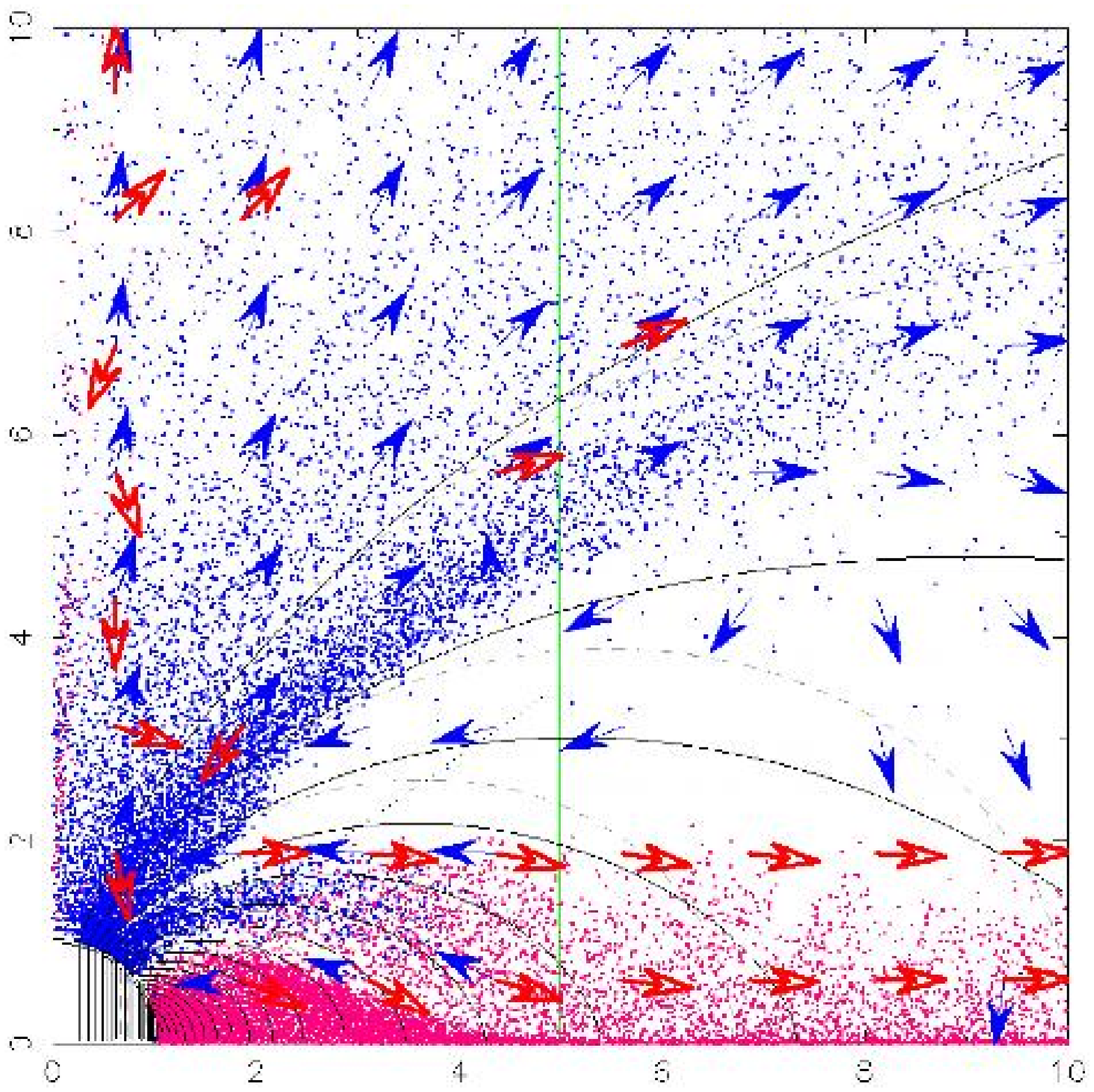} &
     \FigureFile(80mm,80mm){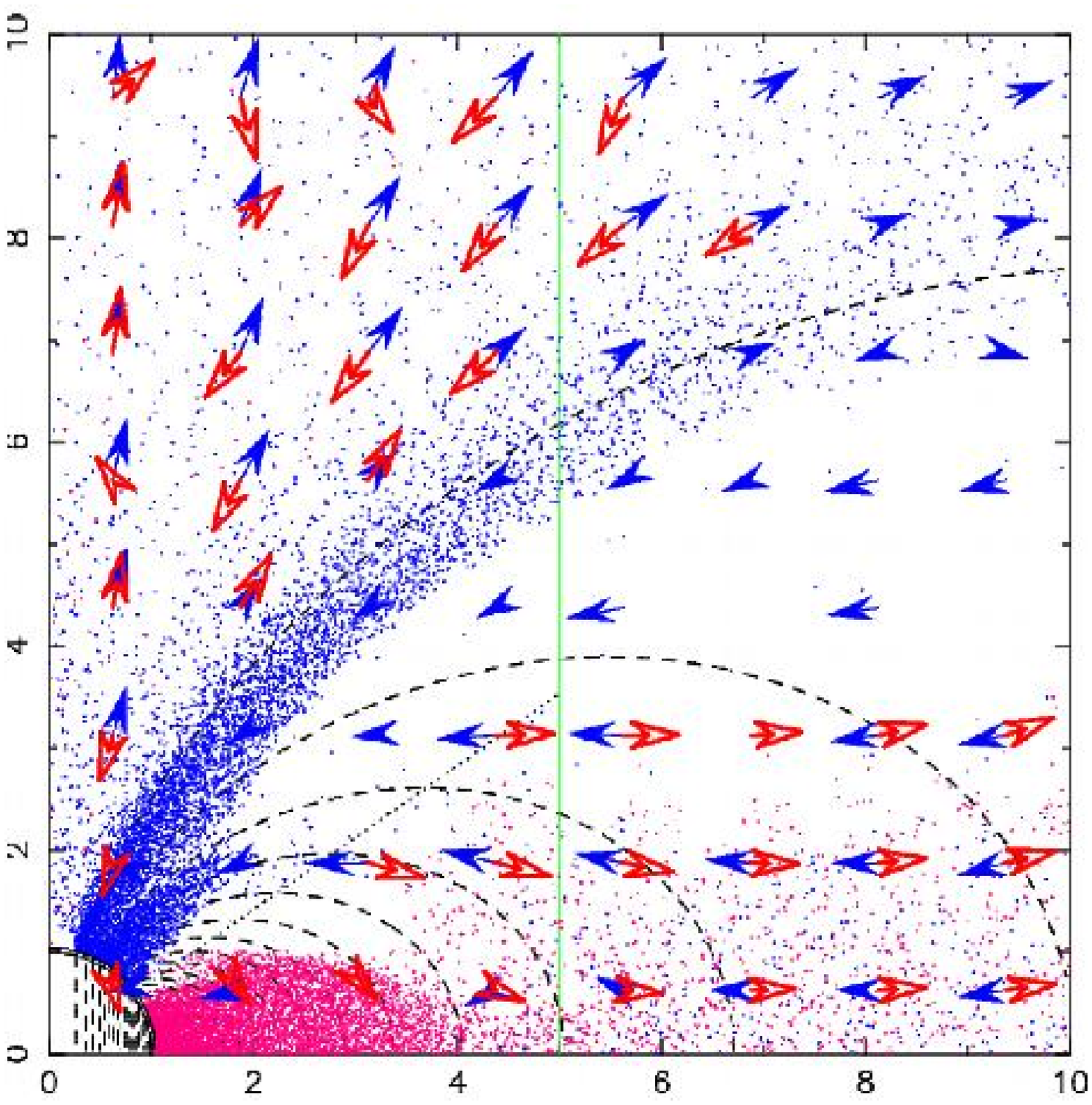} \\
     \FigureFile(80mm,80mm){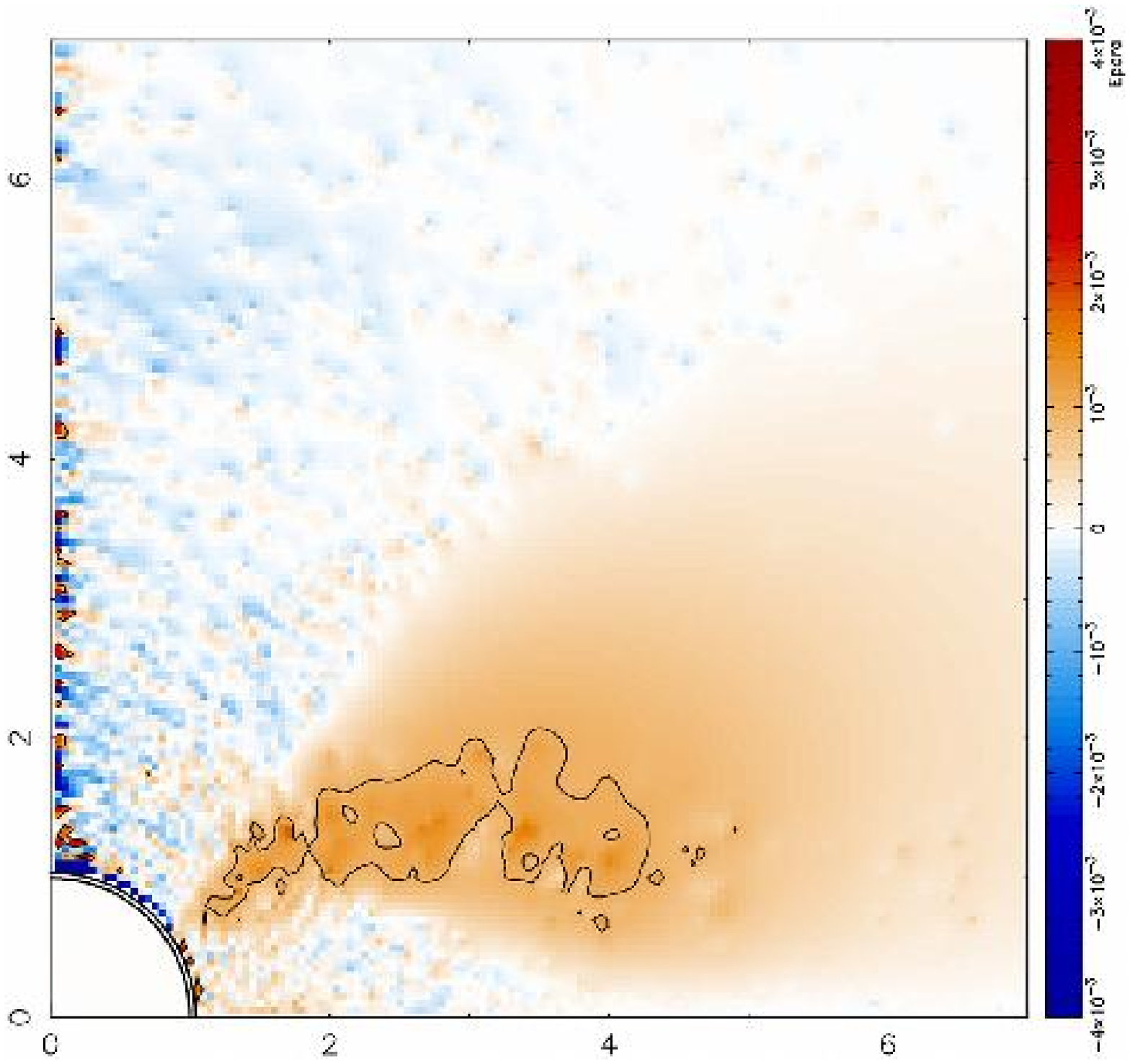} &
	 \FigureFile(80mm,80mm){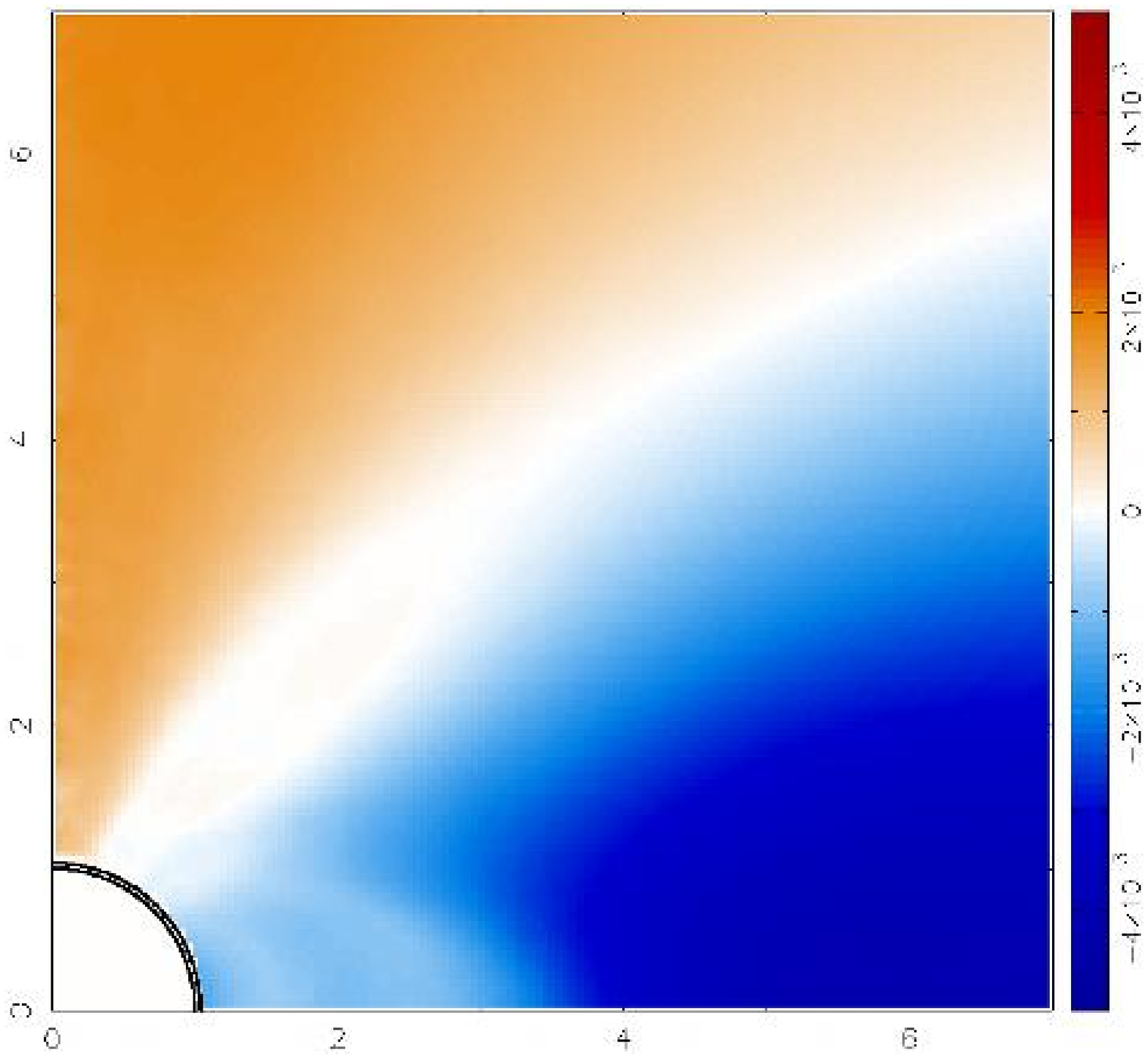}
    \end{tabular}
  \end{center}
    \caption{
   Top left panel:
   Particle distribution (dots) and velocity fields (arrows) on the
   meridional plane in the case of the modified magnetic field.
   Red dots and arrows are for positive particles, and
   blue ones are for negative particles.
   The star is at the bottom left.
   The dotted lines show the dipole magnetic field lines, and the solid
   lines show modified magnetic field lines.
   The geometrical scale length is normalized by the stellar radius, and
   the light cylinder is located at the lateral distance of 5.
   Top right panel: Particle distribution in the case of the dipole magnetic field.
   Bottom left panel: Color contour of the field-aligned electric field
   $E_{||}$. The black line represents where the value of the electric field
   is equal to $E_{\mathrm {cr}}$.
   Bottom right panel: Color contour of the non-corotational potential $\Phi$.
   }
   \label{fig:res01}
\end{figure*}

Magnetospheric plasmas are represented by $\sim 1 \times 10^5$ super-particles.
After integration of $\sim 10$ periods of time, the amount of particles produced at the stellar surface and the gaps, and
the amount of removed particles are balanced,
so that the total particle number becomes constant in the simulation box.
The net charge of the system also converges to a constant value,
and therefore we can consider  the magnetosphere is in a steady state.
The top left panel of figure \ref{fig:res01} shows particle
distribution and velocity field on the meridional plane in the steady state.
Most of the positive and negative particles are located around the equatorial plane
and polar regions, respectively.
In addition, a near vacuum region appears around the middle-altitudes.

The bottom left panel of figure \ref{fig:res01} shows the strength of the field-aligned electric field $E_{||}$.
The strong $E_{||}$, which is $\gtrsim 0.25 B_{\mathrm L}$, exists around the null charge surface and above the poles.
In total $\sim 3.6 \times 10^4$ electron-positron pairs (super-particle pairs) are generated per rotation.

The former region is identified as the OG (see WS).
However, note that the strong $E_{||}$ above the poles still has artificial effects
because it was found that it decreases with increasing the frequency of the particle emission on the stellar surface.
When we changed this emitting frequency from $\sim 45$ times to $\sim 230$ times per rotation,
the $|E_{||}|$ above the poles decreased from $\sim 0.036$ to $\sim 0.016$.
This indicates that if the emission frequency becomes higher, the value of $E_{||}$ would approach $E_{||} \sim 0$.
However, it is difficult to remove such an artificial effect in the global particle simulation due to a limited computational power.
It is not clear at the moment whether the $|E_{||}|$ above the poles includes the essential electric field or not.

The pairs generated in the OG are accelerated in opposite directions by
the $E_{||}$.
The positive particles are accelerated outward, and flowed out as a pulsar wind.
On the other hand, the negative particles are accelerated toward the star by the outward-directed $E_{||}$.
If the star solely absorbs the negative particles continuously, it would
be charged up negatively.
To prevent such charging up, the negative particles are re-emitted from the stellar surface around the magnetic poles and spread into the magnetosphere.
In this way, a current passing through the PC originates the negative particles generated in the OG.

In our simulation, particles are traced up to the distance of $10 R_{\mathrm L}$.
Most of them flow out beyond the outer boundary as a pulsar wind, but some negative particles emitted from the PC region return to the star.
This circulating flow can cross the magnetic field due to the $\boldsymbol{F}_{\mathrm {rad}} \times \boldsymbol{B}$ drift.
In order to understand how modification of the magnetic field lines affects the structure of the magnetosphere,
we have also performed the simulation with the magnetic field fixed to the dipole field.
In the case of the dipole field, the magnetic flux crossing
the equatorial plane beyond $R_{\mathrm L}$ and $10 R_{\mathrm L}$ are
$\Psi_{\mathrm L}=0.1$ and $\Psi_{\mathrm 10L}=0.01$, respectively.
In contrast, in the case of the modified magnetic field, the values are
$\Psi_{\mathrm L} \sim 0.14$ and $\Psi_{\mathrm 10L} \sim 0.03$, respectively.
The closed magnetic fields are pushed out and the
magnetic flux passing through the outer boundary is increased.
This indicates that the open flux is increased.
The top right panel of figure \ref{fig:res01} shows the particle distribution and velocity field in the case of the dipole field.
In comparison with the dipole case, it was found that the circulating flow diminishes in the case of the modified magnetic field.
As for the total number of particles going through the outer boundary,
it is $\sim 3.3 \times 10^4$ per rotation,
which corresponds to $\sim 0.6 \dot{N}_{\mathrm {GJ}}$, where
$\dot{N}_{\mathrm {GJ}} = (\Omega^2 R^3 B_* ) / (2ec)$ is a current of primary particles flowing out from the single polar cap in Goldreich \& Julian model, in the case of the modified magnetic field.

The non-corotational potential $\Phi$ represents deviation of the electric field from the corotational one.
We assume $\Phi = 0$ on the stellar surface.
The bottom right panel of figure \ref{fig:res01} gives the map of
$\Phi$, where if the region is white, then $\Phi \approx 0$
(i.e. the corotation region).
In our results, the polar regions are reddish due to the potential drop just above the stellar surface,
while the low-latitude region, except for the equatorial closed magnetic
flux, is bluish due to the outer gap.

Note that the regions other than the white region in the bottom right panel of figure \ref{fig:res01} do not always
mean $E_{||} \neq 0$.
Even if $\Phi \neq 0$, the field-aligned electric field can be zero when $\boldsymbol{B} \cdot \nabla \Phi = 0$.
On the other hand, the white region where  $\Phi = 0$ is connected with
the star through the magnetic field lines on which  $E_{||} = 0$, and  corotates with the star.

There are two distinguishable corotation white regions.
One is the traditional dead zone with closed magnetic flux (the reason why it is rather small shall be discussed later).
The other is in the middle latitudes extending along the field lines.
This elongated corotation region separates the OG regions with out-going
current from the PC regions with in-going current.
In other words, the middle-latitude corotation region is formed where
the directions of the currents change.
Let us call this region ``the current-neutral zone''.
In our numerical experiment, there is no significant flow of both positive
and negative particles along the magnetic field lines in the current-neutral zone.  
The cloud of negative particles is naturally formed there.
The current-neutral zone is formed in both the dipole and modified magnetic field cases, although the locations were slightly different
because it is formed along the magnetic field lines.

\begin{figure}
  \begin{center}
      \FigureFile(80mm,80mm){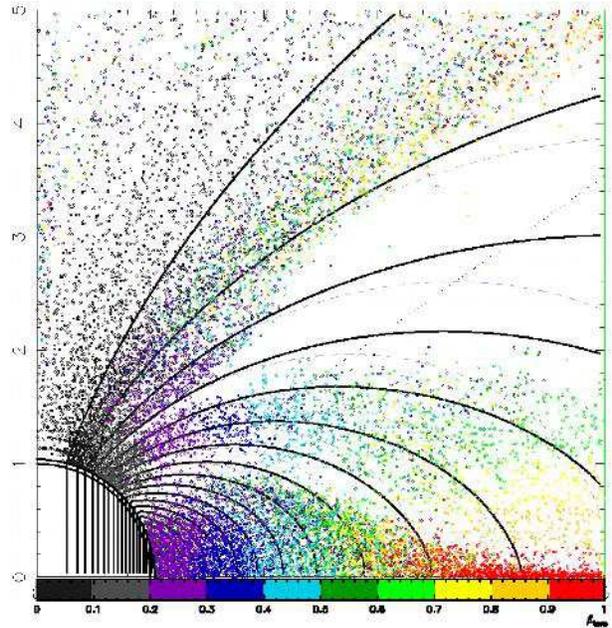}
  \end{center}
    \caption{The color-coded azimuthal velocity of the particles
   (circles). The meaning of the colors is shown in the bar just
   below the equatorial plane, where each color indicates the
   corotational speed at each axial distance.}
   \label{fig:res02}
\end{figure}

\begin{figure}
  \begin{center}
     \FigureFile(80mm,80mm){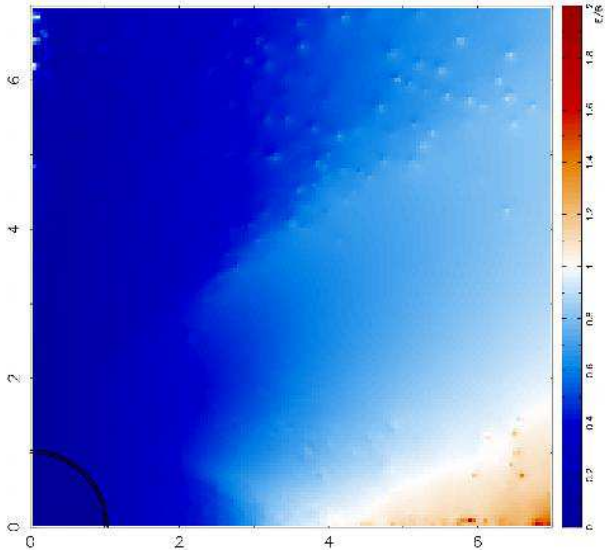}
  \end{center}
\caption{Color contour of the ratio of the magnitude of the electric
   field to the magnitude of the magnetic field,
 $|\boldsymbol{E}|/|\boldsymbol{B}|$.}
   \label{fig:res03}
\end{figure}

The traditional dead zone is defined by the magnetic field lines closed within the light cylinder,
but the one in our simulation does not reach the light cylinder, and is obviously small.
Additionally, the shape of the disk does not follow the magnetic field lines.
As shown in figure \ref{fig:res02}, the inner part of the disk corotates with the star,
but on the other hand the super-rotation, which is caused by the
$E_{||}$ in the gap, took place at the outer part of the disk.
The Lorentz factor increases to saturate by radiation drag, and it is 
$\sim 10$ for the super-particle.
This Lorentz factor would correspond in the actual pulsar to 
 $\sim 10^7 P_1^{-1/4} B_{12}^{1/4} R_6^{3/4}$,
at which the Lorentz force and the radiation drag are equal to each
other near the light cylinder.
For this reason, $\boldsymbol{F}_{\mathrm {rad}} \times \boldsymbol{B}$ drift causes cross-field motion within the light cylinder,
so that the particles leak from the outer part of the disk toward the light cylinder.
The primary force on the particles are the Lorentz force, and therefore
shows gyro-motion,
thus the approximation formula for the radiation drag force can be
applied.
The secondary force is radiation drag and causes the drift motion.
Thus, the size of the corotational disk becomes smaller.
As for the ratio of the magnitude of the electric field to the magnitude of the magnetic field,
the electric field is dominant in the wedge-shaped region beyond the light cylinder (figure \ref{fig:res03}).
This feature, whose apex angle is $\approx 28^\circ$, is similar to the
feature found in the force-free solution
by Uzdensky (2003), whose apex angle is $\approx 62^\circ$.

\section{Discussion and Conclusion}
\begin{figure}
 \begin{center}
 \FigureFile(80mm,80mm){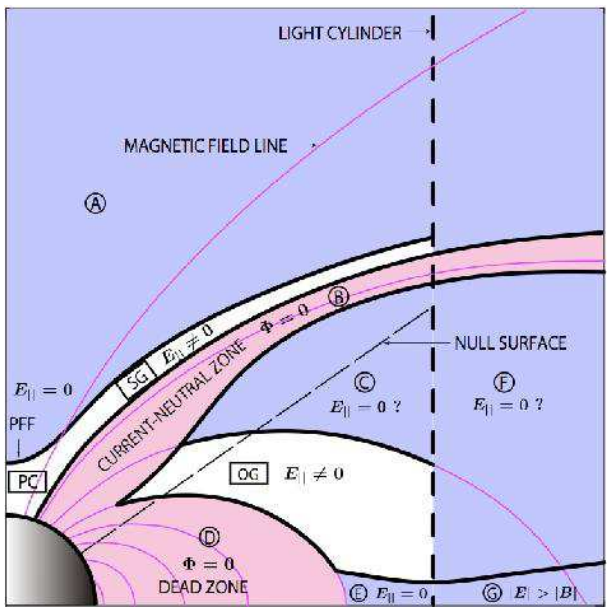}
 \label{fig:dis01}
  \end{center} 
 \caption{
 Geometry of the pulsar magnetosphere on the meridional plane.
 }
\end{figure}
Let us discuss the structure of the global magnetosphere on the basis of our results.
In figure \ref{fig:dis01}, we summarize electromagnetic properties of each magnetospheric region.
There is a possibility that multiple acceleration regions exist in a
different way from the traditional picture because of the current-neutral zone.

First, we discuss Region A and Region B.
In PC and SG models, the side boundary condition is that $\Phi=0$ along the last open field lines.
In our simulation, the side boundary of the polar flows correspond to the rim of Region B.
Therefore, the surface of Region B plays the role of the side boundary of PC and SG models, instead of the traditional dead zone.
The boundary condition on the neutron star surface depends on conditions such as the surface temperature, the surface magnetic field strength, and the atomic number of matter in the surface layer. 
In the case of most pulsars, one may assume $E_{||}=0$ and $\Phi=0$ for the boundary conditions on the stellar surface.
As described in the previous section, the strong $E_{||}$ just above the
magnetic poles (figure \ref{fig:dis01}) includes the artificial effects,
which are caused by the limited emitting frequency of the particles from the stellar surface.
If the emitting frequency is increased, the value of the $E_{||}$ will
likely approach zero.

In space-charge limited flow models, in which $E_{||}=0$ and $\Phi=0$ on
the stellar surface, particles are accelerated and emit gamma-rays.
Then magnetic pair production occurs, and the PFF is formed.
We applied a simple calculation method for generating pairs in order to avoid heavy computations.
However, if we could apply a more accurate and realistic calculation,
the PFF would be properly treated, and the PC and SG would appear in a region bounded by the stellar surface and Region B.
Additionally, we also performed a simulation in which pair creation
was suppressed above the magnetic poles.
Even in this simulation, we obtain an steady state where the OG is formed.
This simulation corresponds to a pair-starved polar cap (PSPC) \citep{b21,b22}, in which the amount of pair creation is not enough to completely screen the accelerating electric field.
It was found in this simulation that the structure of Region B was
almost unchanged as compared with the previous simulation.
This indicates that the current-neutral zone is concrete. In summary,
the acceleration, like polar-slot gap or PSPC, may
exist in a region bounded by Region B.
It is noted that the pair creation in the OG is the key to make the
magnetosphere active even in the aligned rotator,
although the diocotron instability is also proposed to play a similar role (see, eg, \cite{b16,b17}).
If pair creation does not occur anywhere, the dome-disk magnetosphere, such as the magnetosphere in \citet{b18}, is formed in our simulation.
In contrast, if the pair creation occurs in the OG, the generated
electrons go back to the star and are re-emitted around the magnetic poles, and then, they flow out to the magnetosphere.
In this way, pair creation in the OG can change a static magnetosphere into an active magnetosphere without diocotron instability.

Next, let us discuss the region around the OG.
The corotation Region D, where $\Phi=0$, is identified as a traditional dead zone, but Region D is smaller than the traditional one and does not reach the light cylinder.
Instead, Region E exists between Region D and the light cylinder.
In Region E, $E_{||}=0$, but $\Phi \neq 0$ as seen in figure \ref{fig:dis01} because the magnetic field lines which pass through Region E go through the OG, where $E_{||} \neq 0$.
The super-rotation arises in Region E due to the gap.
Note that the super-rotation appears even if $E_{||} = 0$ in the disk. 
This means that rotation speeds of particles approach the speed of light within the light cylinder (see figure \ref{fig:res02}),
and the particles in the region can leak out due to the cross-field motion.
The $E_{||}$ in the OG is $\gtrsim 0.25 B_{\mathrm L}$,
and the potential drop in the OG is $\sim 20\%$ of the effective
electromotive force in the simulation.
In the case of actual young pulsars, the pair creation rate is higher, so that the strength of the $E_{||}$ in the gap should be smaller.
In addition, super-particles can cross the field lines easier due to
their large masses. 
Therefore, it is thought that the size of Region E in the actual
magnetosphere is much smaller than that of the simulation.

\begin{figure}
  \begin{center}
      \FigureFile(80mm,80mm){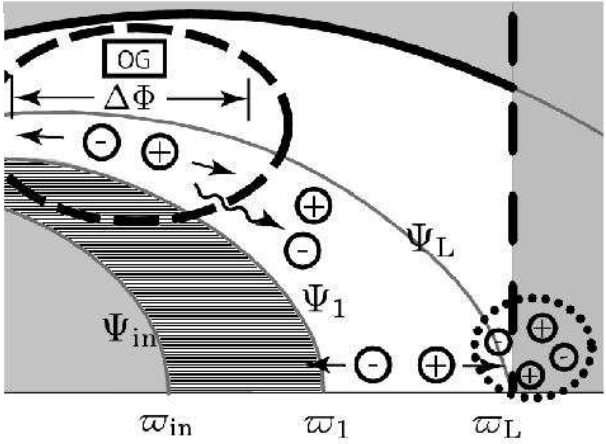}
  \end{center}
    \caption{
   The diagram of the super-rotation region and outer gap.
   The electric potential charges along on the field line $\Psi_{\mathrm
   L}$ due to the outer gap. The potential difference between
   $\Psi_{\mathrm {in}}$ and $\Psi_{\mathrm L}$ becomes larger than the
   corotational one, and therefore a stronger $\boldsymbol{E}_\perp$.
   Thus super-rotation appears.
   }
  \label{fig:dis02}
\end{figure}

The azimuthal velocity in the closed field region originates from trans-field potential difference via $\boldsymbol{E}_\bot \times \boldsymbol{B}$ drift motion,
which may be given by
\begin{equation}
v_\phi = c \varpi \frac{{\mathrm d} \phi}{{\mathrm d} \Psi},
\label{eq:dis02}
\end{equation}
where $\phi$ is the scalar potential.
Let us define the corotating dead zone by $\Psi \le \Psi_{\mathrm {in}}$, and
$\phi_{\mathrm {in}} = (\Omega /c) \Psi_{\mathrm {in}}$ on its surface (see figure \ref{fig:dis02}).
Outside of this corotating dead zone, but within the field line tangent to the light cylinder, denoted by $\Psi_{\mathrm L}$,
we have a super-rotating region with closed field lines ($\Psi_{\mathrm {in}} < \Psi < \Psi_{\mathrm L}$).
The electric potential on the field line $\Psi_{\mathrm L}$ is $\phi_{\mathrm L} = (\Omega/c) \Psi_{\mathrm L}$ on the stellar surface,
but owing to the potential drop $\Delta \Phi$ along the field line in the gap, the potential on the same field line and on the equatorial plane is reduced to $\phi_{\mathrm L} -\Delta \Phi$.
Thus, the potential difference between $\Psi_{\mathrm {in}}$ and $\Psi_{\mathrm L}$ is increased to provide super-rotation:
\begin{eqnarray}
v_\phi & \approx & c \varpi \frac{\phi_{\mathrm {in}} - (\phi_{\mathrm L} - \Delta \Phi)}{\Psi_{\mathrm {in}} - \Psi_{\mathrm L}} \\ 
& = & \varpi \Omega \left(  1 + \frac{c}{\Omega} \frac{\Delta \Phi}{\Psi_{\mathrm {in}} - \Psi_{\mathrm L}} \right) > \varpi \Omega,
\label{eq:dis03}
\end{eqnarray}
which indicates that somewhere within the light cylinder, the rotation
velocity formally exceeds the light speed, say at $\varpi = \varpi_1$.
In reality, radiation drag in the azimuthal direction drives
trans-field motion in a radial direction, where just within $\varpi_1$, and beyond $\varpi_1$. 
In our simulation, the thickness of the super-rotation region is
artificially enlarged because of the somewhat large value of $E_{\mathrm {cr}}$.
The actual thickness of this region $\Delta \varpi = R_{\mathrm L} -
\varpi_1$, can be estimated by (\ref{eq:dis03}).
Providing $\Delta \varpi \sim R_{\mathrm L} (\Psi_{\mathrm {in}} - \Psi_{\mathrm L}) \Psi_{\mathrm L} \ll R_{\mathrm L}$,
we have $\Delta \varpi / R_{\mathrm L} \sim (\Delta \Phi / \phi_{\mathrm L})^{1/2} \approx (\mu/10^{30} {\mathrm G} {\mathrm {cm}}^3)^{1/2} P$,
where $\Delta \Phi \sim 10^{13}$ Volt  is assumed.
In the Crab pulsar, $\Delta \varpi \sim 0.03 R_{\mathrm L}$.
However, in old pulsars, where $\Phi$ becomes comparable to
$\phi_{\mathrm L}$, the super-rotation region must be significant.

There would be no pair creation on the magnetic field flux between $\varpi_1$ and $\varpi_{\mathrm {in}}$, and the inner part of the OG.
If pair creation occurred there, the pairs would not be able to cross
the closed magnetic field lines and accumulate.
If so, it is expected that the inner part of the OG is filled with the pairs, and the gap would disappear.
However, this is inconsistent with the initial assumption that the gap exists.
Thus, the pair-starved region must exist in the inner part of the OG.

In the super-rotational region of the actual pulsar, 
some positrons which generated in the OG flow out at an amount 
approximate to the amount of the GJ current to the outside of the
magnetosphere by the $\boldsymbol{F} \times \boldsymbol{B}$ drift.
On the other hand, some electrons return to the star, and the amount of
this current is thought to be that of the GJ current, at most.
However, the number of pairs generated in the OG is $\sim 100$ times
larger than that of the primary particles.
This indicates that most of the pairs do not become a current.
They might leak as a plasmoid when the particle energy density becomes larger
than that of the magnetic field
(see figure \ref{fig:dis02}).
We later suggest magnetic reconnection at the top of Region E.

Region E mostly consists of positrons.
The positron flow is divided into two kinds of flows in the simulation
with the magnetic field fixed to be dipole.
One is a flow along the equatorial plane to escape from the
magnetosphere, and the other one is a flow circulating the magnetosphere
to return back to the star from the poles.
However, in the case of the modified magnetic field, the circulating component disappears, and most of the positrons flow out as a pulsar wind.
In the present simulation, the mean free path for pair creation is set
at zero for simplification, and the generated pairs do not have initial
outward velocities.
This approximation works well at acceleration regions where pair creation occur actively, but does not work well around the acceleration regions whose structure would not be reproduced correctly in our simulation.
Generated pairs are immediately charge-separated by the
accelerating electric field in the gap, and only positrons can escape
out of the magnetosphere.
In the actual pulsar, the mean free path has a finite value, so that
gamma-rays can cross the field and produce pairs outside the OG.
Thus, not only positrons, but also electron-positron pairs can escape outside the OG.
As pointed out by \citet{b19}, gamma-rays from the gap always propagate in the convex sides of the magnetic field lines,
and subsequently, copious electron-positron pairs must quench the
field-aligned electric field above the OG (Region C and Region F). 
In addition to the trans-field effect, some pairs generated in the outer
part of the OG can escape to Region E and Region F due to the initial outward velocities. 
In this way, Region C, Region E, and Region F should be filled with quasi-neutral plasmas, so that the condition $E_{||}=0$ would be satisfied there. 
If the pair creation rate is made higher than the present value so that the current reaches the GJ value,
some magnetic field lines would be opened.
In that case, the magnetic neutral sheet around the equatorial plane (Region G) would be formed.
The inner edge of the neutral sheet, which is the open-close boundary of the magnetic field structure, is called Y-point.
\citet{b20} studied the Y-point via an axisymmetric particle-in-cell simulation. 
They demonstrated that the magnetic reconnection occurred quasi-periodically and thereby plasma was heated and accelerated.
They also pointed out that the electric-field dominant region ($|\boldsymbol{E}| > |\boldsymbol{B}|$) was formed around the equatorial plane. 
Our simulation treats steady state, so that time-dependent phenomena (eg, the magnetic reconnection) do not appear.
However, Region G also holds the condition $|\boldsymbol{E}| > |\boldsymbol{B}|$ in our simulation (figure \ref{fig:res03}).

As for the directions of the currents, the direction in PC, SG, and PSPC models is inward and the direction in the OG model is outward.
The mutual location of the accelerating regions with opposite currents and how the return current is maintained are troublesome problems. 
One possible picture may be indicated in our simulation.
Note that the inward and outward currents in our simulation exist along the different field lines separated by the current-neutral zone (Region B).

The size of the OG is determined by balance of the amount of supply
and loss of pairs.
We find in our simulation that the size of the OG decreases with increasing 
supply of pairs.
In the actual pulsars, the supply is influenced by the pair creation rate, i.e.,
the gamma-ray emissivity and the mean free path for photon-photon collision to
make pairs.
On the other hand, the loss is due to escaping of pairs from the closed
field region.
However, the loss process is somewhat uncertain, depending on the
structure of the closed-open boundary and on the super-rotation.
The super-rotation is affected by the mass-charge ratio of the super-particle.
As we discussed previously, the actual size of the super-rotation
region and the gap size in the simulation can be artificially
enlarged.
Since the pair creation rate is treated as a model parameter and the
particle loss is merely included quantitatively,
it is at the moment difficult to discuss actual gap size as
function of the pulsar parameters such as the period, field strength
and X-ray flux from the surface.

Some observational results support the aforementioned picture.
According to \citet{b23}, \textit{Fermi}-LAT have detected pulsations from PSR B1509-58 up to 1 GeV with a light curve presenting two
gamma-ray peaks P1 and P2, which was observed at phases $0.96 \pm 0.01$ and $0.33 \pm 0.02$, respectively.
P2 was consistent with the outer magnetosphere geometry although a sharp cutoff was not well explained.
Considering the extension to 1 GeV, and the magnetic pair and photon-splitting attenuation limits, P1, which precedes the radio peak, must also originate in the outer magnetosphere.
However, its phase location could be explained by neither two-pole
caustic (TPC) models (eg, \cite{b24}), which might be realized in SG acceleration models, nor OG models.
PSPC models can produce the gamma-ray peak preceding the radio peak \citep{b8},
so that PSPC emission might be able to explain the phase location of P1.
Thus to explain both peaks, both pair-starved and non-pair-starved (gap)
models are required to co-exist \citep{b23}.
As mentioned, we also performed the simulation without pair creation
above the PC, and found that the structure around the OG was not affected
by eliminating pairs in the PCs.
This indicates that both pair-starved and non-pair-starved gaps exist simultaneously.

\citet{b1} discovered that the distribution of the separation of the two gamma-ray peaks appears to be bimodal with no strong dependence on pulsar $\dot{E}$ (or age).
The $\Delta$ distributions of the separation for each PC, TPC, and OG models are calculated by \citet{b13}.
However, the distribution does not show the bimodal feature, but if two
acceleration regions exist in a magnetosphere, this bimodal feature could be explained.
The bimodal distribution might indicate that we look at different gamma-ray emitting regions from various viewing angles.
From this viewpoint, we can naturally understand the fact that there is no relationship between $\Delta$ and $\dot{E}$.

It is considered that the radio emission from most pulsars come from above the magnetic poles. 
The radio emission is thought to be related to pair creation although the mechanism is still not fully understood.
Therefore, an acceleration region to produce pairs is required above the
magnetic poles.
The possibility that gamma-rays originate from the PC was ruled out by recent observations (eg, \cite{b9,b1}).
Nevertheless, acceleration regions above the magnetic poles, and polar cap cascade are required to explain the radio emission.
Peak separations between radio peaks and gamma-ray peaks indicate that acceleration regions are located at different regions.
These facts also support the idea that at least two kinds of acceleration regions exist in a pulsar magnetosphere.
Explaining both radio and gamma-ray light curves would require both low- and high- altitude acceleration regions.

In conclusion, the above observational facts strongly suggest that both OG and polar-slot gap/PSPC exist.
From our simulation, these acceleration regions are located at above and below the current-neutral zone.
Our present simulation is not enough to resolve the structure of the PC
and SG because the sizes of these regions are insignificant compared to the global magnetosphere.
As a future work, high-accuracy simulations, which have a larger
number of particles, or follow more realistic pair creation processes, are required to discuss the PC and SG in more detail.
Also, it would be challenging to study the magnetospheric structure in the case of oblique rotators.

\section*{Acknowledgments}
Numerical computations were carried out on GRAPE system at Center for Computational Astrophysics, CfCA, of National Astronomical Observatory of Japan.


\begin{thebibliography}{99}
\bibitem[Abdo et al.(2010a)]{b23} Abdo,~A.~A. \etal\ 2010a,
\apj, 714, 927
\bibitem[Abdo et al.(2010b)]{b1} Abdo,~A.~A. \etal\ 2010b,
\apjs, 187, 460
\bibitem[Aliu et al.(2008)]{b9} Aliu,~E. \etal\ 2008,
Sci, 322, 1221
\bibitem[Cheng et al.(1986)]{b19} Cheng,~K.~S., Ho,~C., \& Ruderman,~M. 1986,
\apj, 300, 500
\bibitem[Daugherty \& Harding(1996)]{b2} Daugherty,~J.~K., \& Harding,~A.~K. 1996,
\apj, 458, 278
\bibitem[Dyks \& Rudak(2003)]{b24} Dyks,~J., \& Rudak,~B. 2003,
\apj, 598, 1201
\bibitem[Goldreich \& Julian(1969)]{b12} Goldreich,~P., \& Julian,~W.~H. 1969,
\apj, 157, 869
\bibitem[Harding(2009)]{b25} Harding,~A.~K. 2009, in Neutron Stars and
		Pulsars, ed. Becker,~W. (Berlin and Heidelberg:
		Springer-Verlag) 521
\bibitem[Harding et al.(2008)]{b11} Harding,~A.~K., Stern,~J.~V., Dyks,~J., \& Frackowiak,~M. 2008,
\apj, 680, 1378
\bibitem[Hirotani(2008)]{b6} Hirotani,~K. 2008,
\apj, 688, L25
\bibitem[Krause-Polstorff \& Michel(1985)]{b18} Krause-Polstorff,~J., \& Michel,~F.~C. 1985,
\aap, 144, 72
\bibitem[Makino et al.(2003)]{b14} Makino,~J., Fukushige,~T., Koga,~M., \& Namura,~K. 2003,
\pasj, 55, 1163
\bibitem[Muslimov \& Harding(2003)]{b3} Muslimov,~A.~G., \& Harding,~A.~K. 2003,
\apj, 588, 430
\bibitem[Muslimov \& Harding(2004)]{b4} Muslimov,~A.~G., \& Harding,~A.~K. 2004,
\apj, 606, 1143
\bibitem[Muslimov \& Harding(2004)]{b21} Muslimov,~A.~G., \& Harding,~A.~K. 2004,
\apj, 617, 471
\bibitem[Muslimov \& Harding(2009)]{b22} Muslimov,~A.~G., \& Harding,~A.~K. 2009,
\apj, 692, 140
\bibitem[P\'etri(2009)]{b17} P\'etri,~J. 2009,
\aap, 503, 1
\bibitem[Romani \& Watters(2010)]{b7} Romani,~R.~W., \& Watters,~K.~P. 2010,
\apj, 714, 810
\bibitem[Spitkovsky \& Arons(2002)]{b16}
Spitkovsky,~A., \& Arons,~J. 2002, in Neutron Stars in Supernova Remnants,
ed. Slane,~P.~O., \& Gaensler,~B.~M., (San Francisco: ASP) 81
\bibitem[Takata \& Chang(2007)]{b10} Takata,~J., \& Chang,~H.~K. 2007,
\apj, 670, 677
\bibitem[Takata et al.(2004)]{b5} Takata,~J., Shibata,~S., \& Hirotani,~K. 2004,
\mnras, 354, 1120
\bibitem[Umizaki \& Shibata(2010)]{b20} Umizaki,~M., \& Shibata,~S. 2010,
\pasj, 62, 131
\bibitem[Venter et al.(2009)]{b8} 
Venter,~C., Harding,~A.~K., \& Guillemot,~L. 2009,
\apj, 707, 800
\bibitem[Wada \& Shibata(2007)]{b15} 
Wada,~T., \& Shibata,~S. 2007,
\mnras, 376, 1460
\bibitem[Watters et al.(2009)]{b13} 
Watters,~K.~P., Romani,~R.~W., Weltevrede,~P., \& Johnston,~S. 2009,
\apj, 695, 1289
\end{thebibliography}
\end{document}